\documentstyle[12pt]{article}

\newcommand{\df}{\stackrel{\bigtriangleup}{=}}

\begin{document}

\title {\Large \bf On Bounds  \protect for $E$-capacity of DMC $^*$\footnotetext{$^*$ The author is with the Institute for Informatics and Automation Problems of the Armenian National Academy of Sciences, 1 P. Sevak str., Yerevan 0014, Armenia. E-mail: evhar@ipia.sci.am.}}
\author {\normalsize Evgueni A. Haroutunian, \it Associate Member, IEEE}
\date{}
\maketitle

\noindent {\it Abstract--}Random coding, expurgated and sphere packing bounds are derived by method of types and method of graph decomposition   for $E$-capacity of discrete memoryless channel (DMC). Three decoding rules are considered, the random coding bound is attainable by each of the three rules, but the expurgated bound is achievable only by maximum-likelihood decoding. Sphere packing bound is obtained by very simple combinatorial reasonings of the method of types. The paper joins and reviews the results of previous hard achievable publications.

\vspace{3mm}

\begin{small}

\hspace{1.7 cm} method of types, method of graph decomposition, decoding rules.
\end{small}

\vspace{10mm}

\centerline{\rm \sc I. Introduction}
\vspace{5mm}

Let ${\cal X},{\cal Y}$ be finite sets and $W=\{W(y|x), x\in{\cal X},  \,y\in {\cal Y}\} $ be a stochastic matrix. 

{\it Definition 1:} A {discrete channel} $W$ with {input alphabet}  ${\cal X}$ and {output alphabet} ${\cal Y}$ is defined by stochastic matrix of transition probabilities
$W:{\cal X}\to {\cal Y}.
$
An element $W(y|x)$ of the matrix is a conditional probability of receiving the symbol $y\in {\cal Y}$ on the channel's output  if the symbol $x\in {\cal X}$ is transmitted from the input.

The model for $N$ actions of the channel $W$ is described by the stochastic matrix
$$
W^N:{\cal X}^N\to {\cal Y}^N,
$$ 
the element of which $W^N({\bf y}|{\bf x})$ is a conditional probability of receiving vector ${\bf y}\in {\cal Y}^N$, when ${\bf x}\in {\cal X}^N$ is transmitted. Here we consider memoryless channels, which operate at each moment of time independently of the previous or next transmitted or received symbols, so for all ${\bf x}\in {\cal X}^N$ and ${\bf y}\in {\cal Y}^N$
$$
W^N({\bf y}|{\bf x})=\prod\limits_{n=1}^NW(y_n|x_n).
$$
Let ${\cal M}$ denotes the set of messages to be transmitted and $M$ -- the number of messages. 

{\it Definition 2:} A {code} $(f, g)$ for the channel $W$ is a pair of mappings, where $f:{\cal M}\to {\cal X}^N$ 
is {encoding} and 
$g:{\cal Y}^N\to {\cal M}$ 
is {decoding}. $N$ is called the code length, and $M$ is called the code volume.

{\it Definition 3:} The {probability} of {erroneous transmission}  of the message $m\in {\cal M}$ by the channel using code $(f,g)$ is defined as
\begin{equation}
\label{er}
e(m,f,g,N,W)\df W^N({\cal Y}^N-g^{-1}(m)|f(m))=
\end{equation}
$$
=1-W^N(g^{-1}(m)|f(m)).
$$

We shall consider  
{the maximal probability of error} of the code $(f,g)$:
$$
e(f,g,N,W)\df \max\limits_{m\in {\cal M}}e(m,f,g,N,W),
$$
and the optimal maximal probability of error for the channel $W$:
$$
e(M,N,W)\df \min_{(f,g)} e(f,g,N,W),
$$
where minimum is taken among codes $(f,g)$ of volume $M$,
and the average probability of error for equiprobable messages is
$$
\overline{e}(f,g,N,W)\df \frac 1{M}\sum\limits_{m\in {\cal M}}e(m),
$$
with $\overline{e}(M,N,W)$ as the minimal average probability among all possible codes of the length $N$ and the volume $M$.
It is clear that always
$$
\overline{e}(f,g,N,W)\leq e(f,g,N,W).
$$

{\it Definition 4:} The {transmission rate} of a code $(f,g)$ of volume $M$ is
\begin{equation}
\label{r}
R(f,g,N)\df \frac 1 N \log M.
\end{equation}

Note that in this paper all $\exp$-s and $\log$-s are to the base two.

We consider the codes, error probability of which exponentially decrease with exponent $E$, when $N\to \infty$:
\begin{equation}
\label{11}
e(f,g,N,W)=\exp\{-NE\}.
\end{equation}

Denote  the best volume of the code of length $N$ for channel $W$ satisfying the condition (\ref{11}) for given reliability $E>0$ by $M(E,N,W)$.

{\it Definition 5:} The rate-reliability function, which by analogy with the capacity we call \linebreak  {\bf $E$-capacity}, is for maximal probability of error
$$
R(E,W)=C(E,W)\df \overline{\lim\limits_{N\to \infty}}\frac 1 N \log M(E,N,W),
$$
and ${\overline R}(E, W)$ for the case of average probability of error.
As in the case of capacity, $E$-{\bf capacity} is called {\bf maximal} or {\bf average} and denoted, correspondingly,  $C(E,W)$ or $\overline C(E,W)$ depending on which error probability is considered in (\ref{11}).

The concept of $E$-capacity was first considered by the author in \cite{Hr 67}, where derivation of the upper bound $R_{sp}(E,W)$ was stated. The simple combinatorial proof of $R_{sp}(E,W)$ was obtained in \cite{Hr 82}. In Section 4 for completeness we present it, because the paper \cite{Hr 82} is little-known.

Alternative methods for the existence part of coding theorems demonstration  are Shannon's random coding  and Wolfowitz's maximal code methods. In \cite{CzKr(2) 81} Csisz\'ar and K\"orner introduced a new original method, based on the lemma of Lov\'asz on graph decomposition. Different methods of error exponent investigation were presented in \cite{Db(2) 62} -- \cite{Gl 68} and in many other works. Here we shall derive upper bounds for  $R(E,W)$ using the method of graph decomposition.

\vspace{10mm}

\centerline{\rm \sc II. Formulation of Results}
\vspace{5mm}

In the beginning we remind our notations for necessary characteristics of Shannon's entropy and mutual information and Kullback-Leibler's divergence.

The size of the set ${\cal X}$ is denoted by $|{\cal X}|$. 
Let $P$ be a PD of RV ${X}$ 
$$
P=\{P(x), x\in {\cal X}\},
$$
$V$ be a conditional PD of RV ${Y}$ for given value $x$ of RV ${Y}$
$$
V=\{V(y|x), x\in {\cal X}, y\in {\cal Y}\}.
$$
The joint PD of RVs $X$ and $Y$  is
$$
P\circ V=\{P\circ V(x,y)=P(x)V(y|x), x\in {\cal X}, y\in {\cal Y}\},
$$
and PD of RV $Y$ is
$$
PV=\{PV(y)=\sum\limits_{x\in {\cal X}}P(x)V(y|x), y\in {\cal Y}\}.
$$
Sometimes we need to consider a stocastic matrix $\overline V:{\cal X}\to {\cal X}$ of conditional probabilities $\overline{V} = \{ \overline{V} (x | \tilde{x}), x \in {\cal X}, \tilde{x}\in {\cal X} \}$.

We use the following notations:
for entropy of RV $X$ with PD $P$:
$$
H_P(X)\df -\sum\limits_{x\in {\cal X}}P(x)\log P(x),
$$
for {joint  entropy} of RVs $X$ and $Y$: 
$$
H_{P,V}(X,Y)\df -\sum\limits_{x\in {\cal X},y\in {\cal Y}}P(x)V(y|x)\log P(x)V(y|x),
$$
for {conditional entropy} of RV $Y$ relative to RV $X$:
$$
H_{P,V}(Y|X)\df -\sum\limits_{x\in {\cal X},y\in {\cal Y}}P(x)V(y|x)\log V(y|x),
$$
for {mutual information} of RVs $X$ and $Y$: 
$$
I_{P,V}(X\wedge Y)\df -\sum\limits_{x\in {\cal X},y\in {\cal Y}}P(x)V(y|x)\log \frac{V(y|x)}{PV(y)},
$$
for {informational divergence} of PD $P$ and PD $Q$ on ${\cal X}$:
$$
D(P\Vert Q)\df \sum\limits_{x\in {\cal X}}P(x)\log \frac{P(x)}{Q(x)},
$$
and for {informational conditional divergence} of PD $P\circ V$ and PD $P\circ W$ on ${\cal X}\times {\cal Y}$:
$$
D(V\Vert W|P)\df \sum\limits_{x\in {\cal X},y\in {\cal Y}}P(x)V(y|x)\log \frac{V(y|x)}{W(y|x)}.
$$
The following identities are often useful
$$
D(P\circ V\Vert Q\circ W)=D(P\Vert Q)+D(V\Vert W|P), 
$$
$$
H_{P,V}(X,Y)=H_{P}(X)+H_{P,V}(Y|X)=H_{PV}(Y)+
H_{P,V}(X|Y),
$$
$$
I_{P,V}(X\wedge Y)=H_{PV}(Y)-H_{P,V}(Y|X)=H_{P}(X)+H_{PV}(Y)-H_{P,V}(X,Y).
$$

The proofs in this paper will be based on the method of types \cite{CzKr 81}. The {type $P$ of a sequence} (or vector) ${\bf x}=(x_1,\dots,x_N)\in {\cal X}^N$ is a PD $P=\{P(x)=N(x|{\bf x})/N,\,x\in {\cal X}\}$, where $N(x|{\bf x})$ is the number of repetitions of symbol $x$ among $x_1,\dots,x_N$. 
The {joint type} of ${\bf x}$ and ${\bf y}\in {\cal Y}^N$ is the PD  $P=\{P(x,y)=N(x,y|{\bf x},{\bf y})/N,\,x\in {\cal X},\,y\in {\cal Y}\}$, where
$N(x,y|{\bf x},{\bf y})$ is the number of occurrences of symbols pair $(x,y)$ in the pair of vectors $({\bf x},{\bf y})$.  
We say that the {conditional type} of {\bf y} for given ${\bf x}$ is  PD $V=\{V(y|x),\, x\in {\cal X},\,y\in {\cal Y}\}$ if $N(x,y|{\bf x},{\bf y})=N(x|{\bf x})V(y|x)$ for all $x\in {\cal X},\,y\in {\cal Y}$.

The set of all PD on ${\cal X}$ is denoted by ${\cal P}({\cal X})$ and the subset of ${\cal P}({\cal X})$ consisting of the possible types of sequences ${\bf x}\in {\cal X}^N$ is denoted by ${\cal P}_N({\cal X})$.
The set of vectors ${\bf x}$ 
of type $P$ is denoted by ${\cal T}_P^N(X)$ and ${\cal T}_P^N(X)=\emptyset$   for PD $P\in {\cal P}({\cal X})- {\cal P}_N({\cal X})$.
The set of all sequences ${\bf y}\in {\cal Y}^N$ of conditional type $V$ for given ${\bf x}\in {\cal T}_P^N(X)$ is denoted by ${\cal T}_{P,V}^N(Y|{\bf x})$ and called { $V$-shell} of ${\bf x}$.
The set of all possible $V$-shells for ${\bf x}$ of type $P$ is denoted ${\cal V}_N({\cal Y},P)$.

In the following lemmas   very useful properties of types are formulated, for proofs see \cite{CzKr 81}.

{\it Lemma 1:} (Type counting)
\begin{equation}
\label{l1}
|{\cal P}_N({\cal X})|< (N+1)^{|{\cal X}|},
\end{equation}
\begin{equation}
\label{l1'}
|{\cal V}_N({\cal Y},P)|< (N+1)^{|{\cal X}||{\cal Y}|}.
\end{equation}


{\it Lemma 2:} For any type $P\in {\cal P}_N({\cal X})$
\begin{equation}
\label{l2}
(N+1)^{-|{\cal X}|}\exp\{NH_P(X)\}\leq |{\cal T}_P^N(X)|\leq \exp\{NH_P(X)\},
\end{equation}
and for any conditional type $V$ and ${\bf x}\in {\cal T}_P^N(X)$ 
\begin{equation}
\label{l2'}
(N+1)^{-|{\cal X}||{\cal Y}|}\exp\{NH_{P,V}(Y|X)\}\leq |{\cal T}_{P,V}^N(Y|{\bf x})|\leq \exp\{NH_{P,V}(Y|X)\}.
\end{equation}


{\it Lemma 3:} If ${\bf x}\in {\cal T}_P^N(X)$,  then for every PD $Q$ on ${\cal X}$ 
\begin{equation}
\label{l3}
Q^N({\bf x})=\prod_{n=1}^NQ(x_n)=\exp\{-N(H_P(X)+D(P\Vert Q))\}.
\end{equation}
If ${\bf y}\in {\cal T}_{P,V}^N(Y|{\bf x})$, then for every conditional PD $V$ on ${\cal Y}$ for given ${\bf x}$
\begin{equation}
\label{l3'}
W^N({\bf y}|{\bf x})=\exp\{-N(H_{P,V}(Y|X)+D(V\Vert W|P))\}.
\end{equation}

Consider the  {random coding exponent function} $R_r(E,W)$, which is a lower estimate for $C(E,W)=R(E,W)$

$$
 R_r(P,E,W)\df 
\min\limits_{V:D(V\| W|P)\leq E}\left| I_{P,V}(X\wedge Y) +D(V\| W|P)-E \right| ^{+},
$$
\begin{equation}
\label{n-7'}
{R}_r(E,W)\df \max\limits_PR_r(P,E,W).
\end{equation}

 The {expurgated exponent function} $R_x(E,W)$, which is another lower estimate for $R(E,W)$ is defined using probability matrix $\overline V: {\cal X}\to {\cal X}$
$$
R_x(P,E,W)=\min\limits_{\overline V}\{I_{P,\overline V}(X\wedge \widetilde X)+|{\bf E}_{P,\overline V}d_B(X,\widetilde X)-E|^+\},
$$
here $d_B(x,\widetilde x)$ is the Bhattacharyya distance 
\begin{equation}
\label{Bhat}
d_B(x,\widetilde x)\df -\log \sum_{y\in {\cal Y}}\sqrt{W(y|x)W(y|\widetilde x)},
\end{equation}
 and
$$
R_x(E,W)\df \max\limits_P R_x(P,E,W).
$$

 Sphere packing exponent function serves an upper bound of $R(E,W)$. 
\begin{equation}
\label{Rsp}
{\cal R}_{sp}(P,E,W)=  \min\limits_{V:D(V\Vert W\vert P) \leq E} I_{P,V}(X \land Y), 
\end{equation}
$$
R_{sp}(E,W)=\max\limits_{P}R_{sp}(P,E,W).
$$
It was first considered in \cite{Hr 67}. 

{\it Theorem 1:} For DMC $W$ and for any $E>0$ the following bound holds
$$
R(E,W)\geq \max (R_r(E,W),R_x(E,W)).
$$

{\it Theorem 2:} For DMC $W$ for $E>0$ the following inequalities hold
$$
R(E,W)\leq \overline{R}(E,W)\leq R_{sp}(E,W).
$$

{\it Theorem 3:} For $0<E\leq E_{cr}(P,W)$, where
$$
E_{cr}(P,W)=\min \left\{E:\frac {\partial R_{sp}(P,E,W)}{\partial E}\geq -1\right\},
$$
the estimates are equal each other and give $E$-capacity:
$$
R(P,E,W)=R_{sp}(P,E,W)=R_r(P,E,W). 
$$

For
$$
E_{cr}(W)=\min\left\{E: \frac{\partial R_{sp}(E,W)}{\partial E}\geq -1\right\}=\max\limits_{P} E_{cr}(P,W)
$$
the equality holds:
$$
R(E,W)=R_{sp}(E,W)=R_r(E,W). 
$$

{\it Remark 1}: For  $E\to 0$
$$
\lim_{E\to 0}R_{sp}(P,E,W)=\lim_{E\to 0}R_{r}(P,E,W)=I_{P,W}(X\wedge Y),
$$
$$
R_{sp}(0,W)=R_r(0,W)=C(W).
$$

\vspace{10mm}

\centerline{\rm \sc III. Proof of Theorem 1}
\vspace{5mm}

{\it Lemma 4:}
Consider a finite set ${\cal A}$ and a nonnegative function $\nu$ on ${\cal A}\times{\cal A}$ such that for every $a,b\in {\cal A}$
$$
\nu(a,b)=\nu(b,a),\,\, \nu(a,a)=0.
$$
If for some $t$, for each $ a\in {\cal A}$
$$
\sum_{b\in {\cal A}}\nu(a,b)< t ,
$$
and $t_1,t_2,\dots,t_S$ are nonnegative numbers such that
$$
\sum_{s=1}^St_s\geq t,
$$
then ${\cal A}$ can be divided into $S$ disjoint subsets ${\cal A}_1,\dots,{\cal A}_S$ such that for every $a\in {\cal A}_s$
$$
\sum_{b\in {\cal A}_s}\nu(a,b)< t_s, \,\,s=\overline{1,S}.
$$
 
For proof of the Lemma see \cite{CzKr(2) 81}.
There lower bounds for reliability function $E(R,W)$ of DMC and of sources with side information were obtained using Lemma 4. We now present similar derivation of random coding and expurgated bounds for $E$-capacity $R(E,W)$ of DMC.

Theorem 1 formulated above is a consequence of the following

{\it Theorem 4:}
For DMC $W:{\cal X}\to{\cal Y}$, any $E>0,\,\,\delta>0$ and type $P\in {\cal P}_N({\cal X})$ for sufficiently large $N$ codes $(f,g)$ exist such that 
\begin{equation}
\label{1.1}
e(f,g,N,W)=\exp\{-N(E+\delta)\} 
\end{equation}
and
$$
R(P,f,g,N)\geq \max(R_r(P,E+\delta,W),R_x(P,E+\delta,W)).
$$

The proof of Theorem 4 consists of several steps. First we shall prove

{\it Lemma 5:}
For given type $P\in {\cal P}_N({\cal X})$, for any $0<r<|{\cal T}_P^N(X)|$ such set ${\cal C}$ exists,  that \linebreak  ${\cal C}\subset {\cal T}_P^N(X),\,\, |{\cal C}|\geq r$ and for any $\widetilde{\bf x}\in {\cal C}$ and matrix $\overline V:{\cal X}\to {\cal X}$ different from the identity matrix the following inequality holds
\begin{equation}
\label{1.2}
|{\cal T}_{P,\overline V}^N(X|\widetilde{\bf x})\bigcap {\cal C}|\leq r|{\cal T}_{P,\overline V}^N(X|\widetilde{\bf x})|\exp\{-N(H_P(X)-\delta_N)\},
\end{equation}
where
$$
\delta_N=N^{-1}[(|{\cal X}|^2+|{\cal X}|)\log(N+1)+1].
$$

{\it Proof:} Using  Lemma 4 let us assume ${\cal A}\stackrel{\bigtriangleup}{=}{\cal T}_P^N(X)$ and
$$
\nu({\bf x},\widetilde{\bf x})\stackrel{\bigtriangleup}{=}\left\{
\begin{array}{ll}
|{\cal T}_{P,\overline V}^N(X|\widetilde{\bf x})|^{-1},\,\,\mbox{if}\,\,{\bf x}\neq \widetilde{\bf x} \,\,\mbox{and}\,\,{\bf x}\in {\cal T}_{P,\overline V}^N(X|\widetilde{\bf x}),\\
0,\,\,\mbox{if}\,\,{\bf x}= \widetilde{\bf x}.\\
\end{array}
\right.
$$
Because ${\bf x}$ and $\widetilde{\bf x}$ are of the same type $P$, when ${\bf x}\in {\cal T}_{P,\overline V}^N(X|\widetilde{\bf x})$, then $\widetilde{\bf x}\in {\cal T}_{P,\overline V'}^N(X|{\bf x})$ where $\overline V'$ is such that the matrix with elements $P( \widetilde x)\overline V (x|  \widetilde x)$ is transposed to the matrix with elements $P(x)\overline V' (\widetilde x|x), \,\, x,\widetilde x \in {\cal X}$. Here $NP( \widetilde x)\overline V (x|  \widetilde x)=N(x, \widetilde x | {\bf x}, \widetilde {\bf x})$ and $NP(x)\overline V' (\widetilde x|x)=N(\widetilde x,x | \widetilde {\bf x},{\bf x})$ define correspondingly joint types of the pairs $({\bf x},\widetilde{\bf x})$ and  $(\widetilde{\bf x},{\bf x})$. We have that 
$$
|{\cal T}_{P,\overline V}^N(X|\widetilde{\bf x})|=\prod_{\widetilde x}\frac{(NP( \widetilde x))!}{\prod\limits_{x} (N( \widetilde x,x| \widetilde {\bf x},{\bf x}))!}
$$ 
and
$$
|{\cal T}_{P,\overline V'}^N(X| {\bf x})|=\prod_{ \widetilde x}\frac{(NP( \widetilde x))!}{\prod\limits_{x} (N(x,\widetilde x| {\bf x},\widetilde {\bf x}))!},
$$
the right sides are equal, so we see that $\nu({\bf x},\widetilde{\bf x})=\nu(\widetilde{\bf x},{\bf x})$.
We have also from (\ref{l1'}) 
$$
\sum_{{\bf x}\in {\cal T}_P^N(X)}\nu(\widetilde{\bf x},{\bf x})=\sum_{{\bf x}\in {\cal T}_P^N(X)}\nu({\bf x},\widetilde{\bf x})=\sum_{\overline V}\sum_{{\bf x}\in {\cal T}_{P,\overline V}^N(X|\widetilde{\bf x})}  \nu({\bf x},\widetilde{\bf x})<(N+1)^{|{\cal X}|^2}.
$$
If we take $t\stackrel{\bigtriangleup}{=}(N+1)^{|{\cal X}|^2}$, $t_s\stackrel{\bigtriangleup}{=}t/S,\, s=\overline{1,S}$, then according to Lemma 4 there exists a partition of ${\cal T}_P^N(X)$ into subsets ${\cal A}_s,\, s=\overline{1,S}$, such that for each $\widetilde{\bf x}$ from ${\cal A}_s$
\begin{equation}
\label{1.3}
|{\cal T}_{P,\overline V}^N(X|\widetilde{\bf x})\bigcap {\cal A}_s|\leq \frac 1 S|{\cal T}_{P,\overline V}^N(X|\widetilde{\bf x})|(N+1)^{|{\cal X}|^2},\, s=\overline{1,S}.
\end{equation}
Taking ${\cal C}$ equal to greatest ${\cal A}_s$ and $S$ equal to integer part of $|{\cal T}_P^N(X)|/r$ we receive \linebreak 
$|{\cal C}|\geq S^{-1}|{\cal T}_P^N(X)|\geq r$, and inequality (\ref{1.2}), which follows from (\ref{1.3}) and (\ref{l2}), because 
$$
\frac 1 S=\frac 1 {\left\lfloor { |{\cal T}_P^N(X)| / r}\right\rfloor}\leq \frac 1{|{\cal T}_P^N(X)|/2r}=\frac {2r}{|{\cal T}_P^N(X)|}.
$$
Lemma 5 is proved.

For code existence theorems demonstration it is possible to consider various "good" decoding rules. For definition of those rules following \cite{CzKr(2) 81}, we apply different real-valued functions $\alpha$ defined on ${\cal X}^N\times {\cal Y}^N$. One says that $g_{\alpha}$ decoding is used if to each ${\bf y}$ from ${\cal Y}^N$ on the output of the channel the message $m$ is accepted when codeword ${\bf x}(m)$ minimizes $\alpha({\bf x}(m),{\bf y})$. One uses such functions $\alpha$ which depend only on type $P$ of ${\bf x}$ and conditional type $V$ of ${\bf y}$ for given ${\bf x}$. Such  functions $\alpha$ can be written in the form $\alpha(P,V)$ and at respective decoding  
$$
g_{\alpha}:{\cal Y}^N\to {\cal M},
$$
the message $m'$ corresponds to the vector ${\bf y}$, if
$$
(m',\widetilde V)=\arg\min_{(m,V):\,\, {\bf y}\in {\cal T}_{P,V}^N(Y| f(m)) }\alpha(P, V).
$$
Here $\widetilde V= \{ \widetilde V(y| { \widetilde x}), { \widetilde x} \in {\cal X}, y \in {\cal Y} \}$ is a matrix different from $V$ but guaranteeing that 
$$
\sum\limits_{x \in {\cal X}}P(x)V(y|x)=\sum\limits_{{ \widetilde x} \in {\cal X}}P({\widetilde x}){\widetilde V}(y|\widetilde x), \,\,y \in {\cal Y},
$$
or equivalently, $PV=P{\widetilde V}$.

Previously the following two decoding rules were used \cite {CzKr(2) 81}: {\bf maximum-likelihood decoding}, when the accepted codeword ${\bf x}(m)$ maximizes the transition probability $W^N({\bf y}|{\bf x}(m))$, in this case according to (\ref{l3'})
\begin{equation}
\label{1.4}
\alpha(P,V)=D(V\|W|P)+H_{P,V}(Y|X),
\end{equation}
and the second decoding rule, called {\bf minimum-entropy decoding}, according to which  the codeword ${\bf x}(m)$ minimizing $H_{P,V}(Y|X)$ is accepted, that is
\begin{equation}
\label{1.5}
\alpha(P,V)=H_{P,V}(Y|X).
\end{equation}

In \cite{HrBel} and \cite{HrHr 98} it was proposed another decoding rule by minimization of
\begin{equation}
\label{1.6}
\alpha(P,V)=D(V\|W|P),
\end{equation}
which can be called {\bf minimum - divergence decoding}.

Let $\widetilde {\widetilde V}=\{\widetilde {\widetilde V}(y|x,\widetilde x), x\in {\cal X}, \widetilde x\in {\cal X}, y\in {\cal Y} \},$ be a conditional distribution of $Y$ given values of $X$ and $\widetilde X$ such, that
\begin{equation}
\label{20'}
\sum_{\widetilde x} P(\widetilde x)\overline V( x|\widetilde x)\widetilde {\widetilde V}(y|x,\widetilde x)=P(x)V(y|x), \,\, x\in {\cal X}, y \in {\cal Y},
\end{equation}
\begin{equation}
\label{20''}
\sum_{ x}  P(x)\overline V'(\widetilde x|x)\widetilde{\widetilde V}(y|x,\widetilde x)=P(\widetilde x)\widetilde V(y|\widetilde x), \,\, \widetilde x\in {\cal X}, y \in {\cal Y}.
\end{equation}

Using the notation from \cite{CzKr(2) 81} we write
$$\widetilde V\prec_{\alpha} V \,\, \mbox {if} \,\, \alpha(P,\widetilde V)\leq \alpha(P,V)\,\,\,\mbox{and}\,\,\, P\widetilde V= PV .$$

Let us denote
\begin{equation}
\label{1.7}
R_{\alpha}(P,E,W)\df\min_{{\overline V}, \widetilde {\widetilde V},\widetilde V:\widetilde V\prec_{\alpha} V, V: {D(V\|W|P)\leq E}} \{I_{P,\overline V}(X\land \widetilde X)+|I_{P,\overline V,\widetilde {\widetilde V}}(Y\land \widetilde X|X)+D(V\|W|P)-E|^+\},
\end{equation}
where RV $X,\widetilde X, Y$ have values, correspondingly, on ${\cal X}, {\cal X}, {\cal Y}$ such that the following is valid:

both $X$ and $\widetilde X$ have distribution $P$ and $P\overline V=P$,
 
$\widetilde {\widetilde V}$ is the conditional distribution of RV $Y$ given $X$ and  $\widetilde X$ satisfying (\ref{20'}) and (\ref{20''}).

Minimization in (\ref{1.7}) (and later) is understud by variables ordered from right to left. In (\ref {1.7}) min must be taken by $V$ under condition $D(V\|W|P) \leq E$, then by $\widetilde V$ under condition $\widetilde V\prec_{\alpha} V$, by $\widetilde {\widetilde V}$ under conditions (\ref{20'}), (\ref{20''}) and at last by $\overline V$.

The main portion of Theorem 4 demonstration is contained in

{\it Proposition 1:} For any DMC $W$, any type $P\in {\cal P}_N({\cal X})$, any $E>0,\, \delta_N>0,$ for all sufficiently large $N$ codes $(f,g_{\alpha})$ exist such, that 
\begin{equation}
\label{1.7'}
e(f,g_{\alpha},N,W)= \exp\{-N(E+\delta_N/2)\},
\end{equation}
and the rate is large enough:
\begin{equation}
\label{23}
R(P,f,g_{\alpha},N)\geq R_{\alpha}(P,E+\delta_N,W).
\end{equation}

{\it Proof:}
For some $R$ let us write $r=\exp\{N(R-\delta_N)\}$. According to Lemma 5 for $r<|{\cal T}_P^N(X)|$ a collection ${\cal C}\subset {\cal T}_P^N(X)$ exists such that $|{\cal C}|\geq r$ and 
for each $\widetilde{\bf x}\in {\cal C}$ and any probability matrix $\overline V:{\cal X}\to{\cal X}$ different from the identity matrix,
for $N$ large enough
\begin{equation}
\label{1.8}
|{\cal T}_{P,\overline V}^N(X|\widetilde{\bf x})\bigcap {\cal C}|\leq \exp\{N(R-I_{P,\overline V}(X\wedge \widetilde X))\}.
\end{equation}

Remark that from simmetry we have 
$$
|{\cal T}_{P,\overline V'}^N(\widetilde{X}|{\bf x})\bigcap {\cal C}|\leq \exp\{N(R-I_{P,\overline V'}(\widetilde X \wedge X))\}=
$$
\begin{equation}
\label{e2}
=\exp\{N(R-I_{P,\overline V}(X\wedge \widetilde X))\}.
\end{equation}

 Let us take ${\cal C}$ as a set of codewords of the code $(f, g_{\alpha})$.
If ${\bf x} \in {\cal T}^{N}_{P,\overline V}(X|\widetilde {\bf x})$ exists  such that ${\bf x}\in {\cal C}$ and $\widetilde {\bf x}\in {\cal C}$, then
 \begin{equation}
\label{R}
R(P,f, g_{\alpha}, N)\geq I_{P,\overline V}(X\wedge \widetilde X).
\end{equation}

As in (\ref{er})
$$
e(m,f,g_{\alpha},N,W)=W^N({\cal Y}^N-g^{-1}_{\alpha}(m)|f(m)).
$$
In accordance with $g_{\alpha}$-decoding the set ${\cal Y}^N-g^{-1}_{\alpha}(m)$ contains all words ${\bf y}$ for which codevector $f(m')$ exists different from $f(m)$, such  that,  if  $f(m) \in {\cal T}_{P,\overline V}^{N}(X| f(m'))$ and  \\ ${\bf y} \in {\cal T}_{P,V}^{N}(Y| f(m))\bigcap {\cal T}^N_{P, \widetilde V}(Y|f(m'))$, then the joint type of the triple $(f(m),f(m'),{\bf y})$ is $P\circ \overline V\circ \widetilde{\widetilde V}$, with $\widetilde {\widetilde V}$ meeting conditions (\ref{20'}), (\ref{20''}) and $\widetilde V\prec_{\alpha} V$. 

Denote $({\cal Y}^N-g^{-1}_{\alpha}(m))\bigcap {\cal T}^N_{P, \overline V,  \widetilde {\widetilde V}}(Y|f(m),f(m'))$ the set of such vectors ${\bf y}$, for which the triple $(f(m), f(m'), {\bf y})$ has that joint type $P\circ \overline V\circ \widetilde {\widetilde V}$ for some $m'$ different from $m$. According to (\ref{l1'}) the number of such types does not exceed $(N+1)^{|{\cal X}|^2|{\cal Y}|}$. Taking into account that 
\begin{equation}
\label{x40}
 {\cal T}^N_{P, \overline V,  \widetilde {\widetilde V}}(Y|f(m),f(m'))\subset  {\cal T}^N_{P, V}(Y|f(m)),
\end{equation}
 and then using (\ref{l3'}) we have
$$
W^N({\cal Y}^N-g^{-1}_{\alpha}(m)|f(m))\leq (N+1)^{|{\cal X}|^2|{\cal Y}|}\times
$$
$$
\times \max_{{\overline V}, \widetilde {\widetilde V},\widetilde V:\widetilde V\prec_{\alpha} V, V}  \sum_{f(m'): f(m)\in{\cal T}^N_{P, \overline V}(X|f(m'))\bigcap C}|({\cal Y}^N-g^{-1}_{\alpha}(m))\bigcap {\cal T}^N_{P, \overline V,  \widetilde {\widetilde V}}(Y|f(m),f(m'))|W^N({\bf y}|{f(m)})=
$$
$$
=(N+1)^{|{\cal X}|^2|{\cal Y}|}\max_{{\overline V}, \widetilde {\widetilde V},\widetilde V:\widetilde V\prec_{\alpha} V,V} \sum_{f(m'): f(m)\in{\cal T}^N_{P, \overline V}(X|f(m'))\bigcap C}
|({\cal Y}^N-g^{-1}_{\alpha}(m))\bigcap {\cal T}^N_{P, \overline V,  \widetilde {\widetilde V}}(Y|f(m),f(m'))|\times
$$
\begin{equation}
\label{1.9}
\times \exp\{-N(D(V\|W|P)+H_{P,V}(Y|X))\}.
\end{equation}
Granting (\ref{r}), (\ref{l2'}) and (\ref{1.8}) we bound
$$
\sum_{f(m'): f(m)\in{\cal T}^N_{P, \overline V}(X|f(m'))\bigcap C}|({\cal Y}^N-g^{-1}_{\alpha}(m))\bigcap {\cal T}^N_{P, \overline V,  \widetilde {\widetilde V}}(Y|f(m),f(m'))|\leq
$$
$$
\leq \exp\{N(R(P,f,g_{\alpha},N)-I_{P,\overline V}(X\wedge \widetilde X))\}\exp\{NH_{P, \overline V,  \widetilde {\widetilde V}}(Y|X,\widetilde X)\}=
$$
\begin{equation}
\label{1.10}
=\exp\{N(H_{P,V}(Y|X)-I_{P, \overline V, \widetilde {\widetilde V}}(X,Y\wedge \widetilde X)+R(P,f,g_{\alpha},N))\}.
\end{equation}
Since (\ref{x40}) is valid, from (\ref{l2'}) we have 
$$
\sum_{f(m'): f(m)\in{\cal T}^N_{P, \overline V}(X|f(m'))\bigcap C}|({\cal Y}^N-g^{-1}_{\alpha}(m))\bigcap {\cal T}^N_{P, \overline V,  \widetilde {\widetilde V}}(Y|f(m),f(m'))|\leq
 \exp\{NH_{P,V}(Y|X)\}.
$$
With (\ref{1.10}) it gives us
$$
\sum_{f(m'): f(m)\in{\cal T}^N_{P, \overline V}(X|f(m'))\bigcap C}|({\cal Y}^N-g^{-1}_{\alpha}(m))\bigcap {\cal T}^N_{P, \overline V, \widetilde {\widetilde V}}(Y|f(m),f(m'))|\leq
$$
\begin{equation}
\label{iqx}
\leq 
 \exp\{N(H_{P,V}(Y|X)-|I_{P, \overline V,  \widetilde {\widetilde V}}(X,Y\wedge \widetilde X)-R(P,f,g_{\alpha},N)|^+)\}.
\end{equation}
From (\ref{1.9})  and (\ref{iqx}) we obtain
$$
e(f,g_{\alpha},N,W)=\max_mW^N({\cal Y}^N-g^{-1}_{\alpha}(m)|f(m))\leq 
$$
$$
\leq (N+1)^{|{\cal X}|^2|{\cal Y}|}\max_{{\overline V}, \widetilde {\widetilde V},\widetilde V:\widetilde V\prec_{\alpha} V, V}\exp\{-N(D(V\|W|P)+|I_{P, \overline V,  \widetilde {\widetilde V}}(X,Y\wedge \widetilde X)-R(P,f,g_{\alpha},N)|^+)\}.
$$
If the equality in (\ref{1.7'}) is in force, then for sufficiently large $N$, we have
\begin{equation}
\label{iqx2}
0\geq \min_{{\overline V}, \widetilde {\widetilde V},\widetilde V:\widetilde V\prec_{\alpha} V, V} (D(V\|W|P)+|I_{P, \overline V,  \widetilde {\widetilde V}}(X,Y\wedge \widetilde X)-R(P,f,g_{\alpha},N)|^+-E-\delta_N).
\end{equation}
Suppose that this minimum is obtained on $\widetilde{\widetilde V}_0$ with $V_0$ and $\overline V_0$, that is
$$
0\geq D(V_0\|W|P)-E-\delta_N+|I_{P, \overline V_{0},  \widetilde {\widetilde V}_0}(X,Y\wedge \widetilde X)-R(P,f,g_{\alpha},N)|^+.
$$
Because $|I_{P, \overline V_{0},  \widetilde {\widetilde V}_0}(X,Y\wedge \widetilde X)-R(P,f,g_{\alpha},N)|^+\geq 0$,
we have
$$
0\geq \min_{{\overline V}, \widetilde {\widetilde V},\widetilde V:\widetilde V\prec_{\alpha} V, V: D(V\|W|P)\leq E+\delta_N-|I_{P, \overline V_0,  \widetilde {\widetilde V}_0}(X,Y\wedge \widetilde X)-R(P,f,g_{\alpha},N)|^+}
(D(V\|W|P)-E-\delta_N+
$$
$$
+|I_{P, \overline V,  \widetilde {\widetilde V}}(X,Y\wedge \widetilde X)-R(P,f,g_{\alpha},N)|^+)\geq
$$

$$
\geq \min_{{\overline V}, \widetilde {\widetilde V},\widetilde V:\widetilde V\prec_{\alpha} V, V: D(V\|W|P)\leq E+\delta_N}
(D(V\|W|P)-E-\delta_N+|I_{P, \overline V,  \widetilde {\widetilde V}}(X,Y\wedge \widetilde X)-R(P,f,g_{\alpha},N)|^+).
$$

Since
$$
|I_{P, \overline V, \widetilde {\widetilde V}}(X,Y\wedge \widetilde X)-R(P,f,g_{\alpha},N)|^+\geq I_{P, \overline V,  \widetilde {\widetilde V}}(X,Y\wedge \widetilde X)-R(P,f,g_{\alpha},N),
$$
then we deduce that
\begin{equation}
\label{1.11}
R(P,f,g_{\alpha},N)\geq \min_{{\overline V}, \widetilde {\widetilde V},\widetilde V:\widetilde V\prec_{\alpha} V, V: D(V\|W|P)\leq E+\delta_N}(D(V\|W|P)-E-\delta_N+I_{P, \overline V, \widetilde {\widetilde V}}(X,Y\wedge \widetilde X)).
\end{equation}
From (\ref{1.11}), (\ref{1.7}) and (\ref{R}) we obtain (\ref{23}). Proposition 1 is proved.

{\it Remark 2}: The functions $R_{\alpha}(P,E,W)$ and $R_x(P,E,W)$ depend on $E$ continuously.

{\it Lemma 6:} Let us introduce the following functions
\begin{equation}
\label{1.13}
R_{\alpha,r}(P,E,W)\df\min_{ \widetilde V: \widetilde V\prec_{\alpha} V, V:D(V\|W|P)\leq E}|I_{P,\widetilde V}(Y\wedge \widetilde X)+D(V\|W|P)- E|^+,
\end{equation}
\begin{equation}
\label{1.14}
R_{\alpha,x}(P,E,W)\df\min_{{\overline V}, \widetilde {\widetilde V},\widetilde V:\widetilde V\prec_{\alpha} V, V}\{I_{P,\overline V}(X\wedge \widetilde X)+|I_{P,\overline V,\widetilde {\widetilde V}}(Y\wedge \widetilde X|X)+D(V\|W|P)- E|^+\}.
\end{equation}
Then for all $P$ and $E>0$
$$
R_{\alpha}(P,E,W)\geq \max [R_{\alpha,x}(P,E,W), R_{\alpha,r}(P,E,W)].
$$

{\it Proof:}
The inequality
$$
R_{\alpha}(P,E,W)\geq R_{\alpha,x}(P,E,W)
$$
follows from definitions (\ref{1.7}) and (\ref{1.14}). For the proof of the inequality
$$
R_{\alpha}(P,E,W)\geq R_{\alpha,r}(P,E,W)
$$
remark that
$$
I_{P,\overline V}(X\wedge \widetilde X)+I_{P,\overline V,\widetilde {\widetilde V}}(Y\wedge \widetilde X|X)=I_{P,\overline V,\widetilde {\widetilde V}}(XY\wedge \widetilde X)\geq I_{P,\widetilde V}(Y\wedge \widetilde X)
$$
and then compare (\ref{1.7}) and (\ref{1.13}) using inequality $|a+b|^+\leq |a|^++|b|^+$, which is valid for any real $a$ and $b$.

{\it Lemma 7:}
A point $E^*_{\alpha}(P,W)$ exists, such that
$$
\max [R_{\alpha,x}(P,E,W), R_{\alpha,r}(P,E,W)]=
\left\{
\begin{array}{ll}
R_{\alpha,r}(P,E,W), \,\, \mbox{when}\,\,\, E\leq E^*_{\alpha}(P,W) \\
R_{\alpha,x}(P,E,W), \,\, \mbox{when}\,\,\, E\geq E^*_{\alpha}(P,W).\\
\end{array}
\right.
$$

{\it Proof:} Note that functions $R_{\alpha,r}(P,E,W)$ and $R_{\alpha,x}(P,E,W)$ are nonnegative and decreasing by $E$. Let us first prove that for any $E\geq E'\geq 0$
\begin{equation}
\label{1.15}
R_{\alpha,x}(P,E',W)\leq R_{\alpha,x}(P,E,W)+E-E'.
\end{equation}
In accordance with (\ref{1.14}), bearing in mind the inequality $|a+b|^+\leq |a|^++|b|^+$, we obtain
$$
R_{\alpha,x}(P,E',W)=\min_{{\overline V}, \widetilde {\widetilde V},\widetilde V:\widetilde V\prec_{\alpha} V, V}\{I_{P,\overline V}(X\wedge \widetilde X)+|I_{P,\overline V,\widetilde {\widetilde V}}(Y\wedge \widetilde X|X)+D(V\|W|P)- E'+E-E|^+\}\leq
$$
$$
\leq R_{\alpha,x}(P,E,W)+E-E'.
$$
Denote by $E_{\alpha,r}^0(P,W)$ the least value of $E$, for which $R_{\alpha,r}(E,P,W)=0$. Let us show that for any $E$ and $E'$, such that
$$
0\leq E'\leq E\leq E_{\alpha,r}^0(P,W),
$$
the inequality
\begin{equation}
\label{1.16}
R_{\alpha,r}(P,E,W)+E-E'\leq R_{\alpha,r}(P,E',W)
\end{equation}
holds. Really, in the interval $[0,E_{\alpha,r}^0(P,W))$ function $R_{\alpha,r}(E,P,W)$ is strictly positive, then for such $E$ and $E'$
$$
R_{\alpha,r}(P,E',W)=\min_{ \widetilde V: \widetilde V\prec_{\alpha} V, V:D(V\|W|P)\leq E'}(I_{P,\widetilde V}Y\wedge \widetilde X)+D(V\|W|P)- E)+E-E'\geq
$$
$$
\geq \min_{ \widetilde V: \widetilde V\prec_{\alpha} V, V:D(V\|W|P)\leq E}(I_{P,\widetilde V}(Y\wedge \widetilde X)+D(V\|W|P)- E)+E-E'=
$$
$$
=R_{\alpha,r}(P,E,W)+E-E'.
$$
Denote by $E^*_{\alpha}(P,W)$ the smallest $E$, for which
$$
R_{\alpha,r}(P,E,W)\leq R_{\alpha,x}(P,E,W).
$$
Let us show that this inequality holds for all $E$ greater than $E^*_{\alpha}(P,W)$. Consider two cases. 

If $0\leq E^*_{\alpha}(P,W) < E^0_{\alpha,r}(P,W),$
then it follows from (\ref{1.13}), (\ref{1.14}), (\ref{1.15})  and (\ref{1.16})  that for all $E$ from interval $(E^*_{\alpha}(P,W), E^0_{\alpha,r}(P,W))$
$$
R_{\alpha,r}(P,E,W)+E-E^*_{\alpha}(P,W)   \leq R_{\alpha,r}(P,E^*_{\alpha}(P,W),W)\leq
$$
$$
\leq R_{\alpha,x}(P,E^*_{\alpha}(P,W),W)\leq R_{\alpha,x}(P,E,W)+E-E^*_{\alpha}(P,W).
$$

If $E^0_{\alpha,r}(P,W) \leq E^*_{\alpha}(P,W)$, then for all $E$ greater than $E^*_{\alpha}(P,W)$ we have
$$
R_{\alpha,x}(P,E,W)=0=R_{\alpha,r}(P,E,W).
$$ 
In this case $E^*_{\alpha}(P,W)=E^0_{\alpha,x}(P,W)$.

{\it Lemma 8:}
For each $\alpha$-decoding defined in (\ref{1.4}), (\ref{1.5}),  or (\ref{1.6})
\begin{equation}
\label{1.17}
R_{\alpha,x}(P,E,W)\leq R_{x}(P,E,W),
\end{equation}
moreover, for maximum likelihood decoding given by (\ref{1.4}) the equality holds.

{\it Proof:}
First we prove the inequality (\ref{1.17}). As with (\ref{20'}) we have 
\begin{equation}
\label{1.18}
D(V\| W|P)+I_{P, \overline V, \widetilde {\widetilde V}}(Y\wedge \widetilde X|X)=\sum_{x,\widetilde x,y}
 P(x)\overline V(\widetilde x|x)\widetilde {\widetilde V}(y|x,\widetilde x)\log \frac {\widetilde {\widetilde V}(y|x,\widetilde x)}{W(y|x)},
\end{equation}
and by (\ref{20''})
\begin{equation}
\label{1.19}
D(\widetilde V\| W|P)+I_{P, \overline V, \widetilde {\widetilde V}}(Y\wedge X| \widetilde X)=\sum_{x,\widetilde x,y} P(x)\overline V(\widetilde x|x)\widetilde {\widetilde V}(y|x,\widetilde x)\log \frac {\widetilde {\widetilde V}(y|x,\widetilde x)}{W(y|\widetilde x)},
\end{equation}
the left parts of which are equal when $\widetilde V = V$, then from (\ref{1.14})
$$
R_{\alpha,x}(P,E,W) \leq \min_{\overline V, \widetilde {\widetilde V}, \widetilde V: \widetilde V= V}\{I_{\hat P, \overline{V}}(X\wedge \widetilde X)+
$$
$$
+|(1/2)(I_{P, \overline V, \widetilde {\widetilde V}}(Y\wedge \widetilde X|X)+D(V\|W|P))+(1/2)(I_{P, \overline V, \widetilde {\widetilde V}}(Y\wedge \widetilde X|X)+D(\widetilde V\|W|P))- E|^+\}.
$$
From (\ref{Bhat}), (\ref{1.18}) and (\ref{1.19})  denoting
$$
\widetilde {\widetilde V}_1(y|x,\widetilde x)\df (\exp d_B(x,\widetilde x)) \sqrt {W(y|x) W(y|\widetilde x)}=\frac{\sqrt {W(y|x) W(y|\widetilde x)}}{\sum\limits_{y'}\sqrt {W(y'|x) W(y'|\widetilde x)}}
$$
we have
$$
R_{\alpha,x}(P,E,W)\leq \min_{ \overline V,\widetilde {\widetilde V}}(I_{ P,\overline V}(X\wedge \widetilde X)+
|D(\widetilde {\widetilde V}\|\widetilde {\widetilde V}_1|P)+{\bf E}_{P,\overline V}d_B(X,\widetilde X)- E|^+)=
$$
$$
=\min_{\overline V}(I_{P,\overline V}(X\wedge \widetilde X)+|{\bf E}_{P,\overline V}d_B(X,\widetilde X)- E|^+)=R_{x}(P,E,W).
$$
Let us now prove that in the case of maximum likelihood decoding
\begin{equation}
\label{1.21}
R_{x}(P,E,W)=R_{\alpha,x}(P,E,W).
\end{equation}
From the condition $\widetilde V\prec_{\alpha} V$ and  from (\ref{1.4}) we have 
$$
D(\widetilde V\| W|P)+H_{P,\widetilde V}(Y|\widetilde X)\leq D(V\| W|P)+H_{P,V}(Y|X).
$$
In accordance with (\ref{1.18}), (\ref{1.19}) and the last inequality we can deduce
$$
D(V\| W|P)+I_{P,\overline V, \widetilde {\widetilde V}}(Y\wedge \hat X|X)\geq D(\widetilde{\widetilde V}||{\widetilde{\widetilde V}}_{1}|P\circ\overline V')+{\bf E}_{P,\overline V}d_B(X,\widetilde X)\geq {\bf E}_{P,\overline V}d_B(X,\widetilde X).
$$
Hence
$$
|I_{P,\overline V}(X\wedge \widetilde X)+I_{P,\overline V, \widetilde {\widetilde V}}(Y\wedge \widetilde X|X)+D(V\| W|P)-E|^+\geq
$$
$$
\geq |I_{P,\overline V}(X\wedge \widetilde X)+{\bf E}_{P,\overline V}d_B(X,\widetilde X)-E|^+,
$$
which is equivalent to inverse inequality to (\ref{1.17}) and therefore (\ref{1.21}) holds.

{\it Lemma 9:}
For each $\alpha$-decoding
\begin{equation}
\label{1.22}
R_{\alpha,r}(P,E,W)\leq R_{r}(P,E,W),
\end{equation}
moreover, for

-maximum likelihood decoding,

-minimum entropy decoding,

-minimum divergence decoding

\noindent  the equality holds.

{\it Proof:}
The inequality (\ref{1.22}) is valid because (see (\ref{1.13}) and (\ref{n-7'}))
$$
R_{\alpha,r}(P,E,W)\leq\min_{\widetilde V: \widetilde V =V, V:D(V\|W|P)\leq E}|I_{P,\widetilde V}(Y\wedge \widetilde X)+D(V\|W|P)-E|^+= R_{r}(P,E,W).
$$
For the case of maximal likelihood decoding (\ref{1.4})
$$
\widetilde V\prec_{\alpha} V \Longleftrightarrow \left\{
\begin{array}{ll}
D(\widetilde V \| W|P) +H_{P, \widetilde V}(Y| \widetilde X) \leq D(V \| W|P)+H_{P,V}(Y| X)   , \\
P \widetilde V=PV,\\
\end{array}
\right.
$$
hence
\begin{equation}
\label{2.51}
D(\widetilde V \| W|P)+ I_{P,V}(X \wedge Y) \leq D(V \| W|P)+ I_{P,\widetilde V}( \widetilde X \wedge Y).
\end{equation}
Denote for brevity
$$
{\cal G}\df \{ (V, \widetilde V) : \,\, P \widetilde V=PV, \,\,D(V \| W|P) \leq E \,\,\mbox{and}\,\, ( \ref {2.51}) \,\,\mbox{holds} \}.
$$
Thus, we can write instead of (\ref{1.13})
\begin{equation}
\label{2.52}
R_{\alpha,r}(P,E,W)=\min_{(V, \widetilde V) \in {\cal G}} |I_{P, \widetilde V}( \widetilde X \wedge Y)+ D(V \| W|P)- E|^+ \df\min \{B_1, B_2\}, 
\end{equation}
where 
$$
B_{1}\df \min_{(V, \widetilde V) \in {\cal G}, \widetilde V : D(\widetilde V \| W|P)\leq D(V \| W|P) }|I_{P,\widetilde V}(\widetilde X \wedge Y)+ D(V \| W|P)- E|^+,
$$
$$
B_{2}\df \min_{(V, \widetilde V) \in {\cal G}, \widetilde V : D(\widetilde V \| W|P)\geq D(V \| W|P) }|I_{P,\widetilde V}(\widetilde X \wedge Y)+ D(V \| W|P)- E|^+.
$$
In turn neither  $B_1$ nor $B_2$ are not less than $R_{r}(P, E, W)$. Really, if inequalities (\ref{2.51}) and $D(\widetilde V \| W|P)\geq D(V \| W|P)$ hold simultaneously, then $I_{P,V}(X \wedge Y)\leq I_{P,\widetilde V}(\widetilde X \wedge Y)$. Hence 
$$
B_{2}\geq \min_{(V, \widetilde V) \in {\cal G}, \widetilde V : D(\widetilde V \| W|P)\geq D(V \| W|P) }|I_{P,V}(X \wedge Y)+ D(V \| W|P)- E|^+=
R_{r}(P, E, W).
$$

But
$$
B_{1}\geq \min_{(V, \widetilde V) \in {\cal G}, \widetilde V : D(\widetilde V \| W|P)\leq D(V \| W|P) }|I_{P,\widetilde V}({\widetilde X} \wedge Y)+ D(\widetilde V \| W|P)- E|^+= 
$$
$$
=\min_{ \widetilde V : D(\widetilde V \| W|P)\leq E}|I_{P,\widetilde V}(\widetilde X \wedge Y)+ D(\widetilde V \| W|P)- E|^+=R_{r}(P, E, W).
$$
The two latest inequalities along with  (\ref{1.22}) and (\ref{2.52}) provide the statement of the Lemma for the method of maximum-likelihood decoding.

For the case of minimum-divergence decoding  (\ref{1.6}) 
$$
\widetilde V\prec_{\alpha} V \Longleftrightarrow \{ D(\widetilde V \| W|P)\leq D(V \| W|P), \,\, P \widetilde V=PV \}. 
$$  
Then  according to  (\ref{1.13}) and (\ref{n-7'})
$$
R_{\alpha,r}(P,E,W)\geq \min_{ \widetilde V : D(\widetilde V \| W|P)\leq E}|I_{P,\widetilde V}( X \wedge {\widetilde Y})+D(\widetilde V \| W|P)- E|^+= 
R_{r}(P, E, W).$$

For the case of minimum-entropy decoding from (\ref{1.5})
$$
\widetilde V\prec_{\alpha} V \Longleftrightarrow \{ H_{P, \widetilde V}(Y | \widetilde X)\leq  H_{P, V}(Y | X), PV=P{\widetilde V} \},
$$
which implies $I_{P,\widetilde V}(\widetilde X \wedge Y)\geq I_{P,V}(X \wedge Y)$ from where 
$$
R_{\alpha,r}(P,E,W)\geq\min_{V : D(V \| W|P)\leq E}|I_{P, V}( X \wedge Y)+D(V \| W|P)-E|^+= 
R_{r}(P, E, W).
$$

Thus the proof of Theorem 1 is completed by unification of results of Lemmas 4 -- 9 and Proposition 1.

\vspace{10mm}

\centerline{\rm \sc IV. Proof of Theorem 2}
\vspace{5mm}

Let $E$ and $\delta$ be given such that $E>\delta>0$. Let the code $(f,g)$ of length $N$ be defined, $R$ be the rate of the code and average error probability satisfies the analog of condition  (\ref{11}) for $E-\delta$
$$
\overline e(f,g,N,W) = \exp\{-N(E-\delta)\},
$$ 
which according to Definition 3 is
\begin{equation}
\label{2.14}
\frac1 {M} \sum_{m}
{W}^N\{{\cal Y}^N -g^{-1}(m) | f(m)\}=
\exp\{-N(E-\delta)\}. 
\end{equation}
As the number of messages $M$ may be presented by sum of numbers of codewords of different types
$
M=\sum_P|f({\cal M})\bigcap {\cal T}^N_P(X)|,
$
and the number of all types $P\in {\cal P}_N({\cal X})$ is less than $(N+1)^{|{\cal X}|}$ (see (\ref{l1})), then there exists a "major" type $P^*$ such, that

\begin{equation}
\label{Mik}
  |f({\cal M})\bigcap {\cal T}^N_{P^*}(X)|\geq M 
(N+1)^{-|{\cal X}|}.
\end{equation}
Now for any conditional type $V$ in the left part of (\ref{2.14}) we can consider only codewords of type $P^*$ and the part of output vectors ${\bf y}$ of the conditional type $V$
$$
\sum_{m:f(m) \in {\cal T}^N_{P^*}(X)}\!\!\!\!\!\!\! {W}^N\{{\cal T}^N_{P^*,V}(Y| f(m))
-g^{-1}(m) | f(m)\} \leq M\exp \{-N(E-\delta)\}. 
$$
or
$$
\sum_{m:f(m)\in {\cal T}^N_{P^*}(X)}\!\!\!\!\!\!\! \{|{\cal T}^N_{P^*,V}(Y | f(m))| - |{\cal T}^N_{P^*,V}(Y | f(m))\bigcap g_1^{-1}(m)| \}W^N({\bf y}|{\bf x}) \leq
$$
$$
\leq M\exp \{-N(E-\delta)\}. 
$$
From (\ref{l3'}) we obtain
$$
\sum_{m: f(m)\in {\cal T}^N_{P^*}(X)}\!\!\!\!\!\!\!\!\!\!\!\!\! |{\cal T}^N_{P^*,V}(Y | f(m)) |-
{M\exp\{-N(E-\delta)\}  \over
\exp \{-N(D(V \Vert W | P^*)+H_{P^*,V}(Y \vert X))\}}\! \leq
$$
$$
\leq\sum_{m: f(m)\in {\cal T}^N_{P^*}(X)} |{\cal T}^N_{P^*,V}(Y | f(m))\bigcap g^{-1}(m)| .
$$
It follows from the definition of decoding function $g$ that the sets $g^{-1}(m)$ are disjoint, therefore 
$$
\sum_{m: f(m)\in {\cal T}^N_{P^*}(X)} |{\cal T}^N_{P^*,V}(Y | f(m))\bigcap g^{-1}(m)| \leq |{\cal T}^N_{P^*V}(Y)|.
$$
Then from  (\ref{l2'}) we have
$$
|f({\cal M})\bigcap {\cal T}^N_{P^*}(X)|(N+1)^{-| {\cal X}| |{\cal Y}|}\exp\{NH_{P^*,V}(Y|X)\}-
$$
$$
-M\exp\{N(D(V \Vert W | P^*)+H_{P^*,V}(Y \vert X)-E+\delta)\}\leq
\exp\{NH_{P^*V}(Y)\}.
$$
Taking into account (\ref{Mik}) we come to an estimate:
$$
 M \leq {\exp\{NI_{P^*,V}(X\land Y)\}\over (N+1)^{-| {\cal X}| (|{\cal Y}|+1)} - \exp(N(D(V\Vert W | P^*) - E+\delta))}. 
$$
The right part of this inequality can be minimized by the choice of conditional type $V$, keeping the denominator positive, which takes place for large $N$ when the following inequality holds:
$$
D(V\Vert W | P^*) \leq E-\delta.
$$
 The statement of Theorem 2  follows from the definitions of ${\overline R}(E,W)$ and $R_{sp}(E,W)$ and from the continuity by $E$ of the function $R_{sp}(P,E,W)$ .

Similarly  the same bound in the case of maximal error probability can be proved, but it follows also from the given proof.

{\bf Example.} We shall calculate  $R_{sp}(E,W)$ for the binary symmetric channel (BSC).
Consider BSC $W$ with
$$
{\cal X}=\{0,1\}, \,\,
{\cal Y}=\{0',1'\},
$$
$$
W(0'|1)=W(1'|0)=w_1>0, \,\,\,\,\,\,\,
W(0'|0)=W(1'|1)=w_2>0.
$$ 
Correspondingly, for another BSC $V$ on the same ${\cal X}$ and ${\cal Y}$ we denote
$$
V(0'|1)=V(1'|0)=v_1, \,\,\,\,\,\,\,
V(0'|0)=V(1'|1)=v_2.
$$
It is clear that $w_1+w_2=1,\,\, v_1+v_2=1.$

The maximal value of the mutual information $I_{P,V}(X\wedge Y)$ in the defination of $R_{sp}(E,W)$ comes out on $p^*(0)=p^*(1)=1/2$ because of symmetry of the channel, therefore
$$
I_{P^*,V}(X\wedge Y)=1+v_1\log v_1+v_2\log v_2.
$$ 
The condition $D(V\Vert W | P^*) \leq E$ will take the following form
$$
v_1\log \frac {v_1}{w_1}+v_2\log \frac{v_2}{w_2}\leq E.
$$
So, the problem of extremum with restrictions must be solved (see (\ref{Rsp})):
$$
\left\{
\begin{array}{lll}
-(1+v_1\log v_1+v_2\log v_2)=\max\\ \\
-(v_1\log \frac {v_1}{w_1}+v_2\log \frac{v_2}{w_2}-E)=0\\ \\
v_1+v_2=1.
\end{array}
\right.
$$
Using Kuhn-Takker theorem, we find that $\overline v_1, \overline v_2$ give the solution of the problem if and only if  there exist $\lambda_1>0,\, \lambda_2>0$, satisfying the following conditions

$$
\left\{
\begin{array}{lll}
\frac{\partial}{\partial v_i}(-1-v_1\log v_1-v_2\log v_2)+\lambda_1\frac{\partial}{\partial v_i}(-v_1\log \frac {v_1}{w_1}-v_2\log \frac{v_2}{w_2}+E)+\\ \\
+\lambda_2\frac{\partial}{\partial v_i}(v_1+v_2-1)=0, \,\, i=1,2,\\ \\
\lambda_1(v_1\log \frac {v_1}{w_1}+v_2\log \frac{v_2}{w_2}-E)=0,
\end{array}
\right.
$$
which for $\overline v_1$ and $\overline v_2$ giving maximum are equivalent to
\begin{equation}
\label{exik}
\left\{
\begin{array}{ll}
\log \overline v_i+\log e=-\lambda_1(\log \frac {\overline v_i}{w_i}+\log e +\lambda_2),\, \,\, i=1,2,\\ \\
\overline v_1\log \frac {\overline v_1}{w_1}+\overline v_2\log \frac{\overline v_2}{w_2}=E.
\end{array}
\right.
\end{equation}
Solving the first two equations from (\ref{exik}) we obtain
$$
\overline v_i=w_i^{\frac{\lambda _1}{1+\lambda _1}}\,2^{-\frac{1}{1+\lambda _1}(\lambda _1-\lambda _2+1)\log e},\,\, i=1,2.
$$
Let us denote  $\frac{\lambda _1}{1+\lambda _1}=s$ and remember that $\overline v_1+\overline v_2=1$, then as  functions of parameter  $s\in (0,1)$:
$$
\overline v_1=\frac {w_1^s}{w_1^s+w_2^s},\,\,\,\,\,\,\,
\overline v_2=\frac {w_2^s}{w_1^s+w_2^s}.
$$
From the third condition in (\ref{exik}) we obtain the parametric expressions for $E$ and $R_{sp}(E,W)$:
$$
E(s)=\frac {w_1^s}{w_1^s+w_2^s}\log \frac {w_1^{s-1}}
{w_1^s+w_2^s}+\frac {w_2^s}{w_1^s+w_2^s}\log \frac {w_2^{s-1}}{w_1^s+w_2^s},
$$
$$
R_{sp}(s)=1+\frac {w_1^s}{w_1^s+w_2^s}\log \frac {w_1^{s}}
{w_1^s+w_2^s}+\frac {w_2^s}{w_1^s+w_2^s}\log \frac {w_2^{s}}{w_1^s+w_2^s}.
$$

It is not complicated to see that we arrived to the same relation between $R_{sp}$ and $E$ as that given in Theorem 5.8.3 of the Gallager's book \cite{Gl 68}.

\vspace{10mm}

\centerline{\rm \sc V. Comparison of the Bounds for $E$-capacity}
\vspace{5mm}

{\it Lemma 10:} For given  DMC $W$, for  type $P$ and numbers $0 \leq E' \leq E$
$$
R_{r}(P,E,W)=\min_{E': E'\leq E}|R_{sp}(P, E', W)+E'-E|^+.
$$

{\it Proof:}
Applying definitions (\ref{n-7'}) and (\ref{Rsp}) we see:
$$
R_{r}(P,E,W)=\min_{V : D(V \| W|P)\leq E}|I_{P, V}( X \wedge Y)+D(V \| W|P)-E|^+=
$$
$$
=\min_{E': E'\leq E, V : D(V \| W|P)= E'}|I_{P, V}( X \wedge Y)+E'-E|^+=
$$
$$
=\min_{E': E'\leq E}|R_{sp}(P, E', W)+E'-E|^+.
$$

{\it Lemma 11:}
Involving
$$
E_{cr}(P, W)=\min\{E: \,\, \frac{ \partial R_{sp}(P, E, W)}{ \partial E}\geq-1\}.
$$
we can write for all $E>0$
$$
R_{r}(P,E,W)=\left\{
\begin{array}{ll}
R_{sp}(P, E, W), \,\, \mbox{if}\,\,\, E\leq E_{cr}(P, W), \\
|R_{sp}(P,E_{cr}(P, W), W)+E_{cr}(P, W)-E|^+,\, \mbox{if}\, E\geq E_{cr}(P, W).\\
\end{array}
\right.
$$

{\it Proof:}
Since function $R_{sp}(P, E, W)$ is convex by $E$ then for the values of $E \leq E_{cr}(P, W)$ the slope of the tangent is not greater than $-1$, and for $E\geq E_{cr}(P, W)$, it is equal or greater than $-1$. In other words
$$
\frac{R_{sp}(P, E, W)-R_{sp}(P, E', W)}{E-E'}\leq -1, \,\,\, \mbox{when} \,\, E'<E\leq E_{cr}(P, W),
$$
from where
$$
R_{sp}(P, E, W)+E<R_{sp}(P, E', W)+E',
$$
and consequently
$$
\min_{E': E'\leq E\leq E_{cr}(P, W)}R_{sp}(P, E', W)+ E'=R_{sp}(P, E, W)+E.
$$
We obtain from this equality and Lemma 10 the statement of the Lemma for the case $E\leq E_{cr}(P, W)$. Now, if $E_{cr}(P, W)\leq E'<E$, then
$$
\frac {R_{sp}(P,E,W)-R_{sp}(P,E',W)}{E-E'}\geq -1,
$$
or
$$
R_{sp}(P,E,W)+E\geq R_{sp}(P,E',W)+E',
$$
and consequently
$$
\min\limits_{E':E_{cr}(P, W)\leq E'}R_{sp}(P,E',W)+E'=R_{sp}(P,E_{cr}(P, W),W)+E_{cr}(P, W).
$$
Again, using Lemma 10 and the last equality we obtain that for the case $E\geq E_{cr}(P, W)$
$$
R_r(P,E,W)=\min\{
\min\limits_{E':E_{cr}(P, W)\leq E'<E}|R_{sp}(P,E',W)+E'-E|^+,
$$
$$
\min\limits_{E':E'\leq E_{cr}(P, W)}|R_{sp}(P,E',W)+E'-E|^+\}=
$$
$$
=|R_{sp}(P,E_{cr}(P, W),W)+E_{cr}(P, W)-E|^+.
$$

Thus in the interval $(0,E_{cr}(P,W)]$ the functions $R(P,E,W)$ are exactly determined by 
$$
R(P,E,W)=R_{sp}(P,E,W)=R_r(P,E,W),
$$
and in the interval $(0, E_{cr}(W))$,
$$
R(E,W)=R_{sp}(E,W)=R_r(E,W).
$$
So the proof of Theorem 3 is completed.

\end{document}